\documentstyle[12pt]{article}

\begin{document}
\begin{flushright}
CERN-TH.6604/92
\end{flushright}
\vskip 0.25in
\begin{center}

{\bf \LARGE
Next-to-next-to-leading order QCD analysis of the Gross-Llewellyn Smith
sum rule and the higher twist effects}
\end{center}
\vskip 0.2in
\begin{center} {\bf
Ji\v{r}\'{\i} Ch\'{y}la}
 \\Institute of Physics, Na Slovance 2, Prague 8, Czechoslavakia\\and
\\{\bf Andrei L. Kataev}
 \\Theory Division, CERN, CH-1211 Geneva 23, Switzerland;
\\
 Institute for Nuclear Research, Moscow 117312, Russia\footnote{
Permanent address.}.
\end{center}
\vskip 0.5in
\begin{center}
{\bf \Large Abstract}
\end{center}
\vskip 0.15in
 We present the next-to-next-to-leading order QCD
analysis of the Gross-Llewellyn Smith (GLS)
 sum rule in deep inelastic lepton-nucleon
scattering,  taking into account dimension-two, twist-four power
                                                  correction.
We discuss in detail the renormalization scheme dependence
of the perturbative QCD approximations, propose a procedure for an
approximate
treatment of the quark mass threshold effects and compare
the results of our analysis to the recent experimental data of the
                                                               CCFR
collaboration. From this comparison we extract the value of the strong
coupling constant $\alpha_{s}^{nnl}(M_{Z},\overline{\rm MS})=
0.115\pm0.001(stat)\pm0.005(syst)\pm0.003(twist)\pm0.0005(scheme)$.
We stress the importance of an accurate measurement of the GLS sum rule
and in particular of its $Q^{2}$ dependence.

\vskip 2cm
\noindent   CERN-TH.6604/92\\
\noindent  August 1992
\addtocounter{page}{-1}
\thispagestyle{empty}
\vfill\eject
\pagestyle{empty}
\clearpage\mbox{ }\clearpage
\pagestyle{plain}
\setcounter{page}{1}

\newpage
\setcounter{equation} {0}

{\bf 1.}~~~
In this paper we continue our investigation of the phenomenological
aspects of the available next-to-next-to-leading order (NNLO)
                                                       perturbative
QCD approximations to measurable physical quantities, and in particular
                                                              of
their implications for a precise determination of the strong coupling
constant $\alpha_s$ at different scales (see \cite{Altarelli} for a
                                                              recent
detailed discussion of this point). In our previous publication \cite
                                                              {paper1}
we have discussed in detail the NNLO QCD predictions
 for  the familiar R-ratios in  e$^{+}$e$^{-}$ annihilation
as well as for the $\tau$-lepton decay rate. The inclusion of the
NNLO corrections calculated in \cite{GKL}
 proved to be quite important, in particular for the latter
quantity, as it significantly decreased the sensitivity of QCD
predictions to the well-known renormalization scheme (RS) ambiguity
\cite{paper1}.
This makes the $\tau$-lepton decay rate a  suitable place for testing
 QCD (for  more detailed discussions of this subject see
refs.\cite{Braaten},\cite{Altarelli}).
Using the results of our work \cite{paper1} the RS ambiguities were
considered also in
\cite{stevenson} for the case of the e$^{+}$e$^{-}$ annihilation in the
resonance region and in \cite{Deberder} for the $\tau$-lepton decay
                                                                 ratio.

Besides the above-mentioned two quantities, the NNLO calculations are
available also for some of the deep-inelastic scattering sum rules
\cite{LTV,LV}. The corresponding next-to-leading order (NLO)
                                                          calculations
can be found in \cite{Chet,r1MSB}.
 In this paper we concentrate on the Gross-Llewellyn Smith (GLS) sum
                                                                 rule
and after discussion of various kinds of theoretical ambiguities
                                                          related
to it, present a phenomenological analysis of recent experimental data
                                                               from
the CCFR Collaboration \cite{CCFR}.

{\bf 2.}~~~
The quantity of interest in our case is the non-trivial part $\Delta$
                                                                  of
the GLS sum rule
\begin{equation}
GLS=\frac{1}{2}\int_{0}^{1}dxF^{\bar{\nu}p+\nu p}_{3}(x,Q^{2})
\label{GLS}
\end{equation}
defined as
\begin{equation}
\Delta=(3-GLS)/3
\label{Delta}
\end{equation}
 As $\Delta=\Delta(Q^{2})$ depends, through the
$Q^{2}$ dependence of the structure function $F_{3}$ itself, on $Q^{2}$,
it would be very useful to have this quantity measured in a broad range
                                                                    of
$Q^{2}$ values. Unfortunately the actual situation with the GLS sum
rule is more complicated, for experimental as well as theoretical
                                                              reasons,
and its analysis burdened with several problems related to the
                                                           treatment
of the low $x$ region. The point is that although
$F_{3}^{\nu p},F_{3}^{\bar{\nu}p}$
are functions of $x$ and $Q^{2}$ only, the integral in (\ref{GLS})
                                                               cannot
in practice be evaluated from experimental data at any finite primary
                                                               energy.
Indeed, the recent  experimental analysis of the CCFR
collaboration \cite{CCFR} gives
\begin{equation}
\Delta_{exp}(\langle Q^{2} \rangle = 3 \ {\rm GeV}^{2} ) =
0.167 \pm 0.006(stat) \pm 0.026 (systematic)
\label{data}
\end{equation}

\newpage

At fixed $Q^{2}$ there is always a minimal value of accessible $x$,
determined
by the requirement that the total hadronic energy $W$, related to $x$
                                                                 and
$Q^{2}$ by $$W^{2}=Q^{2}\frac{1-x}{x}+m_{p}^{2}$$ does not exceed $S$,
                                                                 the
total lepton-proton centre of mass energy:
$x_{min}=Q^{2}/(S+Q^{2}-m_{p}^{2})$.
Small values of $x$ thus require small values of the ratio $Q^{2}/S$
                                                               and if
we want to avoid large uncertainties due to extrapolation of
                                                       experimental
data to the low $x$ region, we are forced to include in the
                                                      experimental
determination of (\ref{GLS})
a broad range of $Q^{2}$ values, down to a few GeV$^{2}$.
In this region we encounter the following theoretical complications:
\begin{itemize}
\item \underline{The quark mass thresholds.} The integration over $x$
                                                                in
(\ref{GLS}) means that the corresponding values of $W$ span the whole
region $S>W^{2}>m_{p}^{2}$. As all the NNLO calculations have been
                                                           performed
for massless quarks, we face the question of the appropriate treatment
                                                                of
quark mass thresholds.
 As for a given $Q^{2}$ the quantity determining the effective number
                                                                 of
massless quarks to be used in evaluating (\ref{GLS}) is $W$, we plot
                                                             in Fig.1
as functions of $Q^{2}$ the relative contributions $w^{i}(Q^{2})$ to
                                                             the GLS
sum rule ($i=3,4,5$)
of the three intervals
\begin{equation}
0<x<x_{1},\;\;x_{1}<x<x_{2},\;\;x_{2}<x<1
\label{x123}
\end{equation}
\begin{equation}
x_{i}=\frac{Q^{2}}{Q^{2}+W^{2}_{i}-m_{p}^{2}},\;\;\;i=1,2,\;\;
W_{1}^{2}=4m_{b}^{2},W_{2}^{2}=4m_{c}^{2}
\label{xi}
\end{equation}
in which the effective $n_{f}$ equals 5,4 and 3 respectively
($m_b=4.5\ GeV$,
 $m_c=1.5\ GeV$ are the $b$ and $c$-quark current on-shell masses
\footnote{For $m_c=1.35\ GeV$, used in
the CCFR analysis, our results change insignificantly.}).
The curves in
Fig.1 correspond to three widely used sets of distribution functions:
Duke-Owens set 1 \cite{DO}, DFLM set 2 \cite{DFLM}
 and Morfin-Tung (fit S in $\overline{\rm MS}$) \cite{MT}.
It is obvious that the fractions $w^{i}$ are only weakly sensitive to the
specific choice of quark distribution functions.
In evaluating the fractions $w^{i}$ we have used the parton model formula
$$GLS=\frac{1}{2}\int^{1}_{0}(u_{v}(x)+d_{v}(x))dx$$ where
$u_{v}(x)=u(x)-\overline{u}(x)$ and $d_{v}(x)=d(x)-\overline{d}(x)$ are
the standard valence quark distribution functions. We see that the
                                                              integral
(\ref{GLS}) is dominated by $n_{f}=3$ only for very low $Q^{2}$, while
                                                                   for
$Q^{2}=3$ GeV$^{2}$, which is the average value of $Q^{2}$ in the data of
\cite{CCFR}, $n_{f}=4$ gives the largest contribution, but with
significant admixture of $n_{f}=3,5$. The sizeable $Q^{2}$ dependence of
the fractions
$w^{i}(Q^{2})$ comes predominantly from the explicit $Q^{2}$-dependence
                                                                    of
the
boundary value $x_{i}, i=1,2$ and not from the $Q^{2}$-dependence of
                                                                  the
distribution functions $u_{v}(x,Q^{2}),d_{v}(x,Q^{2})$. Clearly,
the only entirely consistent way of incorporating the flavour threshold
effects would be to carry out all the calculations up to the NNLO
for massive quarks. As this is practically impossible to do, we shall
                                                                 later
formulate a parton model based procedure which we believe should
                                                             reasonably
approximate the effects of quark mass thresholds.
\item \underline{Renormalization scheme dependence.} For $Q^{2}$ in the
region of a few GeV$^{2}$ the RS ambiguities are expected to be
phenomenologically important, as was the case in \cite{paper1}.
                                                          Moreover, for
the GLS sum rule the RS dependence in the low $Q^{2}$ region turns out
                                                                   to
depend sensitively on the value of $n_{f}$.
\item \underline{Contribution of higher twists.} Contrary to the case
                                                                 of the
$\tau$-lepton decay, which is also characterized by a rather low value
                                                                  of
the natural scale $Q^{2}$, the results \cite{Jaffe,Braun} give
concrete estimates of the dimension-two, twist-four contribution
 (for  recent theoretical discussions of the dimension-two
 contributions to the e$^{+}$e$^{-}$ annihilation and the
 $\tau$-lepton decay R-ratios see \cite{Zakharov,Altarelli}). The
                                                             estimates
\cite{Braun} suggest that the twist-four contribution is quite  sizeable
and has therefore to be taken into account.
\end{itemize}

{\bf 3.}~~~
The estimate of the corresponding twist-four contribution to the GLS
                                                                 sum
rule, performed in \cite{Braun} using QCD sum rules formalism, implies
                                                                   for
our quantity $\Delta$
\begin{equation}
\Delta_{twist-4}=\frac{8}{27}\frac{\langle \langle O^{S}\rangle
                                                             \rangle}
{Q^{2}},\;\;\;
\langle \langle O^{S}\rangle \rangle =0.33 {\rm GeV}^{2}
\label{condensate}
\end{equation}
Although the typical QCD sum rules accuracy is about $30\%$, we prefer
                                                                 to be
conservative and threfore put a $50\%$ error bar on the above estimate.
These errors are consistent with the estimate of the higher twist
contribution
obtained in \cite{Braun} by two other methods, namely vector dominance
approximation and the non-relativistic quark model.
In \cite{Braun} it was also shown that the results of the latter method
indicate the problems in the related bag model
calculations of \cite{Fajfer}, which gave  negligibly small values of
the twist-four contribution.

For $Q^{2}=3$ GeV$^{2}$ we
thus use
\begin{equation}
\Delta_{twist-4}=0.032 \pm 0.016
\label{twist}
\end{equation}
Its central value is about 1/3 of
the leading perturbative correction in the
 $\overline{\rm MS}$ RS.The best way to
detect the presence of the higher twist contributions would be to study
                                                                    the
$Q^{2}$ evolution of the GLS sum rules. Experimentally this is,
                                                           however,
difficult to do and all data available so far therefore correspond to
                                                                  the
averages over rather broad intervals of $Q^{2}$.

{\bf 4.}~~~
The perturbative part of $\Delta(Q^{2})$ can be expanded in the
renormalized couplant (we adopt the notation of \cite{PMS1})
 $a({\rm RS})=\alpha_s({\rm RS})/\pi$
\begin{equation}
\Delta_{pert}(Q^{2})=a({\rm RS})(1+r_{1}(Q^{2},{\rm RS})a({\rm RS})
+r_{2}(Q^{2},{\rm RS})a^{2}({\rm RS})+\cdots)
\label{delta}
\end{equation}
defined in a particular RS.
As indicated
 in (2) both the couplant $a$ and the coefficients $r_{k}$ do
depend on the chosen RS. For the discussion of the
RS dependence of physical quantities like (2), the RS may be uniquely
                                                               defined
by the set $\{a,c_{k} ;k\geq2\}$
where $c_{k}$ are related to the coefficients of the QCD
$\beta$-function. Let us write the couplant $a({\rm RS})$ as
a function of the renormalization scale variable $\mu$,
\begin{equation}
\frac{da(\mu,RC)}{d\ln\mu}=\beta(a)=-ba^{2}(1+ca+c_{2}a^{2}+\cdots)
\label{RG}
\end{equation}
where $b=(33-2n_{f})/6$, $c=(153-19n_{f})/(66-4n_{f})$ are
the RG invariants while
$c_{i},i>2$ are free parameters defining the so-called renormalization
convention (RC). At the NNLO we have two free parameters labelling
our RS:
$c_{2}$ and either $a$ itself or $\mu$, related to it in (\ref{RG}).
                                                                 The
consistency conditions lead to the following explicit relations
                                                            \cite{PMS1}
\begin{equation}
                 r_{1} = b\ln(\mu/\overline{\Lambda})-\rho
\label{con1}
\end{equation}
\begin{equation}
                 r_{2} = \rho_{2} -c_{2} +(r_{1} +c/2)^{2}
\label{con2}
\end{equation}
where
\begin{equation}
b\ln(\mu/\overline{\Lambda})=
\frac{1}{a}+c\ln\left(\frac{ca}{\sqrt{1+ca+c_{2}a^{2}}}\right)+
                                                         f(a,c_{2})
\label{bln}
\end{equation}
and
\begin{eqnarray}
f(a,c_{2}) & = & \frac{2c_{2}-c^{2}}{d}\left(\arctan \frac{2c_{2}a+c}
                                                                {d}
-\arctan \frac{c}{d} \right), \;\;\; d=\sqrt{4c_{2}-c^{2}},\:
                                                       4c_{2}>c^{2}\\
& = & \frac{2c_{2}-c^{2}}{2d}\left(\ln\left|\frac{2c_{2}a+c-d}{2c_{2}a
+c+d}\right|
-\ln \left| \frac{c-d}{c+d}\right|\right), \;\;\;
 d=\sqrt{c^{2}-4c_{2}},\: 4c_{2}<c^{2}\;\;\;
\label{f}
\end{eqnarray}
In (\ref{con1},\ref{con2}) $\rho=\rho(\sqrt{Q^{2}}/\overline{\Lambda})$
and $\rho_{2}$ are RG invariants \cite{PMS1}
and $\overline{\Lambda}$ is defined as
\begin{equation}
\overline{\Lambda}=\Lambda(2c/b)^{-c/b}
\label{def}
\end{equation}
where $\Lambda$, which is held fixed, is the conventional definition
of the QCD $\Lambda$-parameter.
 Combining (\ref{con1},\ref{bln}) we find
\begin{equation}
r_{1}=\frac{1}{a}+c\ln \left(\frac{ca}{\sqrt{1+ca+c_{2}a^{2}}}\right)
+f(a,c_{2})-\rho
\label{R1}
\end{equation}
and putting all together we obtain $\Delta^{nnl}_{pert}$ as a function
                                                                   of
$a,c_{2},\rho$
and $\rho_{2}$. Note that the energy dependence of $\Delta^{nnl}_
                                                             {pert}$
enters entirely through the RG invariant $\rho$ which can be
written as
\begin{equation}
\rho=b\ln (\sqrt{Q^{2}}/\overline{\Lambda}_{\overline{\rm MS}})
-r_{1}(\mu=\sqrt{Q^{2}},\overline{\rm MS})
\label{rho}
\end{equation}
where we take for the referential RS the $\overline{\rm MS}$ one.

  The RS dependence of $\Delta^{nnl}_{pert}(a,c_{2};\rho,\rho_
{2})$ can therefore be represented by a two-dimensional surface in
                                                               three
dimensions. In this picture
each point on such a surface represents uniquely one RS. Recall that at
 the NLO $\Delta^{nl}_{pert}$ was given simply as \cite{PMS1}
\begin{equation}
    \Delta^{nl}_{pert} =a(2+ca\ln (ca/(1+ca))-\rho a)
\end{equation}
and the corresponding curve was close to a parabola.
At the NNLO
  the surface representing $\Delta^{nnl}$ depends non-trivially on the
 mutual relation of the two RG invariants $\rho$ and $\rho_{2}$, and in
particular on the sign of the latter one.
For the  e$^{+}$e$^{-}$ annihilation and $\tau$ decay R-ratios,
 $\rho_{2}<0$ for all $n_{f}\geq 3$ and so only this case was
discussed in
\cite{paper1}. For (\ref{GLS}) the situation is more complicated as now
                                                                    both
cases $\rho_{2}<c^{2}/4$ and $\rho_{2}>c^{2}/4$ are physically relevant.

In \cite{LV} the NNLO calculations of (\ref{GLS}) were carried out in
                                                                  the
$\overline{\mbox{\rm MS}}$ RS with the result
\begin{equation}
r_{2}=41.441-8.02n_{f}+0.177n_{f}^{2}
\label{r2MSB}
\end{equation}
which coupled with the earlier known formulae for $r_{1}$ \cite{r1MSB},
and $c_{2}$ \cite{c2MSB} in the
same $\overline{\mbox{\rm MS}}$ RS
\begin{equation}
r_{1}=55/12-n_{f}/3
\label{r1MSB}
\end{equation}
\begin{equation}
c_{2}=\frac{77139-15099n_{f}+325n_{f}^{2}}{9504-576n_{f}}
\label{c2MSB}
\end{equation}
yields an explicit dependence of $\rho_{2}(n_{f})$ \footnote{Here we
                                                           included
in $\rho_{2}$ the contributions of the light-by-light-type graphs
contributing to $r_{2}$. For the phenomenologically interesting case
$\rho>0$ the separation of these contributions in accordance with the
proposal of \cite{Kataev} has an entirely negligible effect on the
                                                              results
of this analysis.}. In particular
\begin{equation}
\rho_{2}(3)=3.438, \;\;\; \rho_{2}(4)=-0.928, \;\;\;\rho_{2}(5)=-5.351
\label{345}
\end{equation}

To get a quantitative idea of the shape of $\Delta_{pert}^{nnl}$ as
a function of $a$ and $c_{2}$, for given $\rho$ and $\rho_{2}$, we
                                                            look for
the stationary points
with respect to the variation of $a$, given  by  the
solutions to the equation
\begin{equation}
\frac{d\Delta^{nnl}_{pert}}{da}=0
\label{saddle}
\end{equation}
In \cite{paper1} we discussed in detail the situation for
                                                 $\rho_{2}<0$; here
we briefly sketch what it looks like for $\rho_{2}>0$. In this case
the solutions of (\ref{saddle}) make up a curve in the plane
                                                      $a,c_{2}$ and
lie in the quadrant $a>0,c_{2}>0$.
For sufficiently large $\rho$, there lies along each such curve a
                                                            saddle
point defining the PMS choice of the RS. For large values of the
                                                          couplant,
relevant for the IR region, the functional form of the curves, i.e.
$a(c_{2},\rho,\rho_{2})$, can be expressed analytically.
Expanding (\ref{R1})
in powers of $1/a$ and keeping only the non-vanishing terms we get,
                                                              using
(\ref{con1},\ref{con2})
\begin{equation}
\Delta^{nnl}_{pert}
=\left (\rho_{2}-c_{2}+(\gamma-\rho)^{2}\right)a^{3}+
\left(\gamma-\frac{c}{2}
-\rho\right)a^{2}+a+2(\gamma-\rho)\kappa+O(1/a)
\label{rnnl}
\end{equation}
where $\kappa=1/(3c_{2})$ and
\begin{equation}
\gamma=c\ln (c\sqrt{c_{2}})+\frac{2c_{2}-c^{2}}{\sqrt{4c_{2}-c^{2}}}
\left(\frac{\pi}{2}-\arctan \frac{c}{\sqrt{4c_{2}-c^{2}}}\right)+
                                                        \frac{c}{2}
\label{gamma}
\end{equation}
The equation (\ref{saddle}) has two simple physical solutions
\begin{equation}
a_{1}=
\frac{\rho-\gamma+c/2+\sqrt{3(c_{2}-\rho_{2})+
(\rho+c/2-\gamma)^2-3(\rho-\gamma)^{2}}}{3(\rho-\gamma+
                                        \sqrt{c_{2}-\rho_{2}})
(\rho-\gamma-\sqrt{c_{2}-\rho_{2}})}
\label{a1}
\end{equation}
\begin{equation}
a_{2}=\frac{\rho-\gamma+c/2-\sqrt{3(c_{2}-\rho_{2})+
                                         (\rho+c/2-\gamma)^{2}-
3(\rho-\gamma)^{2}}}{3(\rho-\gamma
+\sqrt{c_{2}-\rho_{2}})(\rho-\gamma-\sqrt{c_{2}-\rho_{2}})}
\label{a2}
\end{equation}
which coincide for $c_{2}^{0}$ given as the solution to the
equation
\begin{equation}
3(c_{2}-\rho_{2})+(\rho+c/2-\gamma)^{2}-3(\rho-\gamma)^{2}=0
\label{1}
\end{equation}
At that point
\begin{equation}
a_{1}=a_{2}=\frac{1}{\rho-\gamma+c/2} \Rightarrow\infty\;\;\;as\;\;
\rho\Rightarrow(\gamma-c/2)
\label{a1a2}
\end{equation}
For $\rho<(\gamma-c/2)$ only $a_{2}$ stays positive and is thus
physically acceptable. As we decrease $\rho$ even further the
curves move upwards and simultaneously shrink to a point at some
$\overline{c_{2}}$. The lower bound on $\rho$ follows again from
the requirement $a_{2}>0$ which means
\begin{equation}
\rho>h(c_{2})=\gamma(c_{2})-\sqrt{c_{2}-\rho_{2}}
\label{21}
\end{equation}
The minimum of the function $h(c_{2})$ lies at $\overline{c_{2}}$
                                                            given as
the solution of the
equation
\begin{equation}
-c\sqrt{4c_{2}-c^{2}}+2c_{2}\left(\pi-2\arctan\frac{c}{\sqrt{4c_{2}-
                                                             c^{2}}}
\right)
-\frac{\sqrt{4c_{2}-c^{2}}(4c_{2}-c^{2})}{2\sqrt{c_{2}-\rho_{2}}}=0
\label{22}
\end{equation}
Substituting the solution of (\ref{22}) into (\ref{21}) we get the
lower bound $\rho_{min}^{PMS}$ as a function of $\rho_{2}$. For
$\rho<\rho_{min}^{PMS}$, $\Delta^{nnl}_{pert}$ is again a monotonous
                                                            function
of the couplant for any $c_{2}$. Solving (\ref{22}) for $n_{f}=3$
                                                         we obtain
\begin{equation}
\rho_{min}^{PMS}=0.863,\;\;\overline{c_{2}}=
\lim_{\rho\rightarrow\rho_{min}}c_{2}^{PMS}(\rho)=12.29
\label{min}
\end{equation}

It should be stressed that the basic idea of
 the method of the ``effective charges''(EC) \cite{Grunberg}
(or the scheme-invariant perturbation theory \cite{DG})
should be considered more carefully
 when we go from the NLO to the
NNLO. Indeed the condition
\begin{equation}
r^{EC}=a^{EC}
\label{EC}
\end{equation}
implies at the NNLO merely
\begin{equation}
r_{1}+r_{2}a=0
\label{ECgen}
\end{equation}
which has an infinite number of solutions corresponding to the
                                                       intersection
of
the surface $\Delta^{nnl}_{pert}(a,c_{2})$ with the plane $\Delta=a$.
                                                              There
are two, one or no intersections for any given $c_{2}$ , depending on
                                                                the
values of $\rho$ and $\rho_{2}$. At first glance there is no obvious
                                                             reason
to single out one of them. However, requiring $r_{1},r_{2}$ to vanish
separately (as is assumed in \cite{Grunberg}) fixes $c_{2}=
                                                   \rho_{2}+(c/2)$ as
well as $a$. In the following this is what we call the EC result. The
equation determining the value $\rho_{min}^{EC}$ at which
$a_{EC}^{nnl}\rightarrow\infty$ reads
\begin{equation}
\rho_{min}^{EC}=c\ln\frac{c}{\sqrt{\rho_{2}+c^{2}/4}}+
                                            \frac{4\rho_{2}-c^{2}}
{4\sqrt{\rho_{2}}}\left(\frac{\pi}{2}-\arctan\frac{c}{2\sqrt
                                                    {\rho_{2}}}\right)
\label{rhoechmin}
\end{equation}
and has for $n_{f}=3$ the solution
 $$\rho_{min}^{EC}(3)=1.35$$
At the NNLO and for $n_{f}=4$ the IR fixed points of the PMS and
                                                           EC NNLO
approximants are at
\begin{equation}
a^{*PMS}=1.12,\;\;c_{2}^{*PMS}=-2.18,\;\; a^{*EC}=5.17\;\;c_{2}^{*EC}=
-0.335
\label{IR4}
\end{equation}
while for $n_{f}=5$
\begin{equation}
a^{*PMS}=0.371,\;\;c_{2}^{*PMS}=-10.62,\;\; a^{*EC}=0.594
\;\;c_{2}^{*EC}=-4.95
\label{IR5}
\end{equation}
We shall not advocate here any of the popular choices of the RS
                                                        (PMS, EC or
$\overline{\rm MS}$) but shall take the difference between the PMS/EC
 and the $\overline{\rm MS}$ results as a measure of the RS-dependence
(in the phenomenologically relevant region of $\rho$ the EC results are
practically indistinguishable from those of the PMS).
As we shall see, typical values of $\rho$ appropriate
to the CCFR data \cite{CCFR} lie in the region $\rho\in(2,6)$, while
$\rho$ up to 25 might be of interest at currently accessible values of
$Q^{2}$.

{\bf 5.}~~~
The main difference between the RS dependences for $n_{f}=3,4,5$
                                                          lies in the
IR region, which is most of all influenced by the fact
                                                  that $\rho_{2}<0$
 for $n_{f}=4,5$ and $\rho_{2}>0$  for $n_{f}=3$ (see (\ref{345}))
implying no IR stability at the NNLO for $n_{f}=3$ even
                                                    in the PMS and EC
approaches.
The question to what extent is the IR stability,
                                           observed for $n_{f}=4,5$
in certain schemes, is of physical relevance or merely an artefact
 of finite order calculations and/or the choice of the RS, is
                                                          difficult
to answer on the basis of the available
                                  perturbative calculations themselves.
Nevertheless, if it should really be the case, then a similar behaviour
would certainly
have to be observed for $n_{f}=3$ as well. As this is not the case the
                                                                IR
stability
 of the NNLO PMS and EC approximants for $n_{f}=4,5$, characterized by
(\ref{IR4}),(\ref{IR5}), has probably little physical relevance.
 Moreover, due to a rather small magnitude of $\rho_{2}$,
$\Delta^{nnl}_{pert}$ has IR zeros of the PMS and EC approximants
at so large values of the couplant that the NNLO approximations can
                                                            hardly be
trusted. In the following we shall discuss only the
                                            region of positive $\rho$
where the problem of asymptotic explosion does not arise.

In Fig.2 we plot the dependence $\Delta_{pert}^{nnl}(\rho)$
on $\rho$ for all
three RS: PMS/EC and $\overline{\rm MS}$ and for $n_{f}=3,4,5$. For
  PMS/EC results we observe sizeable $n_{f}$ dependence in particular in
the small $\rho$ region. This difference is significant up to
                                                         $\rho\sim5$.
On the other hand the $\overline{\rm MS}$ results are nearly $n_{f}$
independent!. This somewhat surprising effect is a result of nontrivial
partial compensation between
significant $n_{f}$-dependences of the coefficients
$r_{k},k=1,2$, as given in (\ref{r1MSB},\ref{r2MSB}), and the
                                                      couplant $a$ in
the $\overline{\rm MS}$ RS, induced by the $n_{f}$-dependence of the
$\beta$-function coefficients $b_{k}$.

The same facts, but viewed differently, are presented in Fig.3 where
we plot the RS dependence, as measured by the difference
                                                  between the PMS/EC
and $\overline{\rm MS}$ results, for $n_{f}=3,4,5$.
                                          The PMS and EC approaches
are practically indistinguishable in our plots. At the NNLO and for
fixed $\rho$ the RS dependence diminishes as $n_{f}$ grows, while at
                                                                 the
NLO it remains about the same.
The importance of the NNLO corrections with respect to the
                                                    NLO ones can be
assessed
 from Fig.4, where we plot the comparison between NLO and NNLO as a
function of $\rho$ for all three values if $n_{f}$ and for
                                                   both PMS/EC and
$\overline{\rm MS}$ RS. We see that
\begin{itemize}
\item in the $\overline{\rm MS}$\ RS we find $\Delta^{nl}_{pert}
(\rho)<\Delta^{nnl}_{pert}(\rho)$ for all
values of $\rho$,
 the difference $\Delta^{nnl}_{pert}-\Delta^{nl}_{pert}$
being a decreasing function of $n_{f}$;
\item in the PMS/EC approaches the difference
$\Delta^{nl}_{pert}-\Delta^{nnl}_{pert}$ is
positive for $n_{f}=5$, close to zero for $n_{f}=4$ and negative for
$n_{f}=3$. Moreover this difference is sizeable and thus of
phenomenological
relevance. For the consequences of  these two observations
see further discussion;
\item in the phenomenologically important region of $Q^{2}$
the incorporation of the NNLO corrections into the analysis
decreases the difference between the PMS/EC and $\overline{\rm MS}$
results.
\end{itemize}
The
 curves shown in Figs. 2,3,4 can be used for straightforward
                                                       determination
of the QCD parameter $\Lambda_{\overline{\rm MS}}$
                                          from experimental data on
$\Delta$. For easy use we have fitted them by an analytical
                                                    expression of
the form
\begin{equation}
\Delta_{pert}^{(i)}(\rho)=
\sum_{j=0}^{5}\frac{r_{j}^{(i)}}{\rho^{j}}
\label{fit}
\end{equation}
in the wide interval $\rho\in(2,26)$, with $i$ labelling one of the
                                                                18
combinations of the order (NLO or NNLO), RS (PMS, EC or
                                                $\overline{\rm MS}$)
and $n_{f}$ (3,4,5).
The values of all the parameters $r_{j}^{(i)}$ are given in Table 1.
                                                               The
extraction of $\Lambda_{\overline{\rm MS}}$ from a given
                                                 experimental value
$\Delta_{exp}$ proceeds in two steps: first one solves the equation
$\Delta_{pert}^{(i)}(\rho)=\Delta_{exp}$ and then uses the solution
$\rho_{exp}^{(i)}$ in the formula
\begin{equation}
\Lambda_{\overline{\rm MS}}^{(i)}=
Q\exp \left[-\left(\frac{r_{1}(\overline{\rm
MS})+\rho_{exp}^{(i)})}{b}\right)\right]
 \left( \frac{2c}{b} \right)^{c/b}
\label{extraction}
\end{equation}

The analytical parametrization of the dependences
                                         $\Delta^{(i)}_{pert}(\rho)$
is useful also for the determination of the error
                                         $\sigma_{\Lambda}$ of the
extracted $\Lambda_{\overline{\rm MS}}$ from the error
                                                $\sigma_{\Delta}$ on
$\Delta_{exp}$:
\begin{equation}
\frac{\sigma_{\Lambda}}{\Lambda_{\overline{\rm MS}}}=
\frac{\sigma_{\Delta}}{b\left|d\Delta(\rho)/d\rho\right|}
\label{error}
\end{equation}

The relation (\ref{extraction}) holds for massless quarks and
                                                          thus makes
good sense only provided $n_{f}$ is well defined and fixed. As
 in practice this is usually not the case, we have
                                              developed a procedure
which takes into account at least in an approximate way
                                                   the quark mass
thresholds. For all four combinations of the order (NLO and NNLO)
                                                               and RS
(EC/PMS and $\overline{\rm MS}$, EC and PMS being indistinguishable
                                                              for our
purposes), it consists of the following steps:
\begin{enumerate}
\item From the information on valence distribution functions
                                                         we determine,
for a given $Q^{2}$, the fractions $w^{i}(Q^{2})$  defining
                                                        the relative
importance
of the contribution of $n_{f}=3,4,5$ to (\ref{GLS}). We do this
                                                            by taking
the average of results corresponding to the three
                                             mentioned sets of quark
distribution functions. For $Q^{2}=3$ GeV$^{2}$ we find: $w^{1}=0.20,
w^{2}=0.51, w^{3}=0.29$.
\item From (\ref{extraction}) $\Lambda_{\overline{\rm MS}}^{(n_{f})}$
 appropriate for $n_{f}$ massless quarks is determined.
\item Using the formulae from \cite{Marciano}
these values are then translated to
 $\Lambda_{\overline{\rm MS}}^{(4)}(n_{f})$
   where the argument $n_{f}$ keeps track of the source of this
                                                         resulting
   $\Lambda_{\overline{\rm MS}}^{(4)}$.
\item The average value $\tilde{\Lambda}_{\overline{\rm
 MS}}^{(4)}=\sum^{5}_{i=3}w^{i}\Lambda_{\overline{\rm MS}}^{(4)}(i)$
 similarly for errors) is evaluated.
\item Using the average value $\tilde{\Lambda}_{\overline{\rm MS}}^
                                                             {(4)}$
we go back and calculate the corresponding values of
$\Lambda_{\overline{\rm MS}}^{(n_{f})}$.
\item From these values of $\Lambda_{\overline{\rm MS}}^{(n_{f})}$
 we calculate, according to (\ref{rho}), the corresponding
$\rho(n_{f})$ and then finally evaluate the averages
 $\overline{\Delta}_{pert}=\sum^{5}_{i=3}
w^{i}\Delta_{pert}(\rho_{i})$.
\end{enumerate}

For this procedure to be self-consistent we should come at the end
                                                              of step
6 close to the experimental value of $\Delta_{exp}$ \ref{data}.
If the data do not correspond to a fixed value of $Q^{2}$
the whole procedure should be folded with the known $Q^{2}$
                                                   dependence of the
data. As this is difficult to do we have carried out
                                              the above procedure for
fixed
 $Q^{2}=3$ GeV$^2$, equal to the average value of $Q^{2}$ in
                                                         the data of
\cite{CCFR}.
We have included the higher twist contribution as given in
                                                       (\ref{twist})
 \footnote{We neglected in our analysis the kinematical power
corrections  and the twist-six corrections
since it was shown in ref.\cite{Braun} that they are
significantly smaller than the central value
of the twist-four contributions (\ref{twist}). It might be
interesting to take these effects into account in order
 to compare their value  with the
assumed by us $50\%$ error bars of the twist-four contibutions.}
but
carried out the whole analysis also for the case of no higher
                                                        twists at all.
In Table 2 we present  in detail the results of our
procedure for the case of the central value of $\Delta_{twist-4}=0.032$.
Notice that the averages in the last column are indeed quite close to
$\Delta_{exp}-0.032=0.135$.
Similar tables have been constructed for the upper (0.048) as well as
                                                                 lower
(0.016) estimates of twist-4 contributions and also for the case of no
twist-4 contribution at all. In these latter cases only the
                                                     extracted values
of $\Lambda_{\overline{\rm MS}}^{(4)}$ are presented in Table 3.
{}From  Tables 2 and 3 we conclude that:
\begin{itemize}
\item the inclusion of the higher twists is quite important as it
                                                           lowers the
central value of $\Lambda_{\overline{\rm MS}}$ by about 100 MeV;
\item in the $\overline{\rm MS}$ as well as PMS/EC approaches
$\Lambda_{\overline{\rm MS}}^{(4)}({\rm NNLO},n_{f}=3)<
\Lambda_{\overline{\rm MS}}^{(4)}({\rm NLO},n_{f}=3)$ while in
                                                           the PMS/EC
approach
they are almost the same (smaller) for $n_{f}=4$ ($n_{f}=5$).
This is a consequence of the first two observations of Sec.5,
namely, that $\Delta^{nl}_{pert}<\Delta^{nnl}_{pert}$ in the
$\overline{\rm MS}$ RS and for PMS/EC with $n_f$=3, while in
the latter case for $n_f$=4 ($n_f$=5) we have $\Delta^{nl}_{pert}
\approx \Delta^{nnl}_{pert}$ ($\Delta^{nl}_{pert}>\Delta^{nnl}_{pert}$)
(see Fig.4);
\item the inclusion of the NNLO corrections reduces substantially the RS
dependence as measured by the difference between the results for PMS/EC
and $\overline{\rm MS}$ approaches while its influence on
$\Lambda_{\overline{\rm MS}}$ is
 smaller than the effects of the twist-four corrections;
                                                                  due to
\item the errors of the extracted $\Lambda_{\overline{\rm MS}}^{(4)}$
values are dominated on one side by the systematical experimental
uncertainties
 and on the other by uncertainty in the higher twist contribution.
The reduction of the systematical experimental errors (which are the
biggest single source of errors and make about half of the total error),
 to the level of statistical ones would
make the GLS sum rules a very good place for testing perturbative
QCD provided a more accurate estimate
 of the higher twist were to be  available.
But even taking into account the combined effect of all discussed
                                                              errors
the accuracy of the extracted $\Lambda_{\overline{\rm MS}}$ is not bad;
\item the results of the averaging procedure are not far from those
which correspond to fixed $n_{f}=4$. This is due to the fact that for
$Q^{2}=3$ GeV$^{2}$
$n_{f}=4$ is dominant while $n_{f}=3,5$ contribute with
                                                  comparable strength,
thus roughly balancing each other. For other
                                         $Q^{2}$ the situation may be
quite different.
\end{itemize}

{\bf 6.}~~~Let us now summarize our results. At the NLO and in the
$\overline{\rm MS}$ RS we find
\begin{equation}
\Lambda^{(4)}_{\overline{\rm MS}}=317 \pm 23(stat) \pm 99(syst)
\pm62(twist) \ {\rm MeV}
\label{msbarnlo}
\end{equation}
while the NLO PMC/EC result reads
\begin{equation}
\Lambda^{(4)}_{\overline{\rm MS}}= 241 \pm 14(stat) \pm 60(syst)
 \pm40(twist) \ {\rm MeV}
\label{pmsnlo}
\end{equation}
where the third error, due to higher twist contributions corresponds to
the limits given in (\ref{twist}).
Notice that at the NLO
the value of $\Lambda^{(4)}_{\overline{\rm MS}}$ extracted
within the PMS/EC approach is smaller than the one obtained in
the $\overline{\rm MS}$ analysis. A similar situation occurs
in the NLO analysis of the BCDMS data on F$_2(x,Q^2)$ without higher
                                                                 twist
contributions
\cite{Vovk} and in the NLO analysis of the BCDMS and SLAC data
with higher twist contributions \cite{Kotikov}.
Our analysis differs from that of \cite{Kotikov} and other
                                                      similar ones in
the way higher twists are treated. While we use concrete estimates
for twist-four contributions (though with sizeable errors), previously
they were simply
 parametrized by the corresponding free parameters which were
fitted together with pure perturbative expressions
from experimental data.

At the NNLO the difference between the $\overline{\rm MS}$ and
                                                            the PMS/EC
results becomes considerably smaller:
\begin{eqnarray}
\overline{\rm MS}: \;\;
\Lambda^{(4)}_{\overline{\rm MS}} & = & 265\pm 18(stat) \pm 80(syst)
\pm 50 (twist) \ {\rm MeV}\\
{\rm PMS/EC}: \;\;
\Lambda^{(4)}_{\overline{\rm MS}} & = & 249\pm 16(stat) \pm 70(syst)
\pm45(twist) \ {\rm MeV}
\label{lambdas}
\end{eqnarray}
For a careful reader we add the following comment concerning the
                                                             different
relation between the errors at the NLO and the NNLO in PMS/EC and
$\overline{\rm MS}$ RS. As follows from Figs. 2,3,4, in the PMS/EC
approaches
the central NNLO value of $\Lambda^{(4)}_{\overline{\rm MS}}$
                                                           is slightly
larger than the NLO one. Furthermore, for $n_{f}$=4 the derivative
of the PMS/EC approximant to $\Delta(\rho)$ is somewhat
                                                     bigger at the NLO
than at the NNLO. This leads (see (\ref{error})) to slightly lower
                                                                error
$\sigma_{\Lambda}$ at the NLO. The combination of these two
small effects results in a slight increase of both theoretical
and experimental uncertainties of $\Lambda^{(4)}_{\overline{\rm MS}}$
at the NNLO in PMS/EC approach compared to the corresponding
                                                          NLO analysis
(compare (43) with (41)).

Combining the $\overline{\rm MS}$ and PMS/EC results we finally get
\begin{eqnarray}
\Lambda^{(4)}_{\overline{\rm MS}}({\rm NLO\;}) & = &
279 \pm19  (stat)\pm80  (syst)\pm50  (twist)\pm38  (scheme)
\ {\rm MeV} \\
\Lambda^{(4)}_{\overline{\rm MS}}({\rm NNLO}) & = &
257\pm17  (stat)\pm75  (syst)\pm47  (twist)\pm8   (scheme) \ {\rm MeV}
\label{final}
\end{eqnarray}
Notice that with respect to the NLO results the NNLO statistical,
systematical and higher twist uncertainties are slightly smaller, while
the error due to the RS-dependence is reduced significantly. This is
                                                                the main
effect of including the NNLO corrections in the analysis of
                                                       GLS sum rule. Our
results are in very good
 agreement with  those of the detailed NLO analysis \cite{BCDMS}
(carried out in the
 $\overline{\rm MS}$ RS and including the phenomenological
parametrization of the higher-twists effects)
\footnote{The recent NLO $\overline{\rm MS}$ and
scheme-invariant fits \cite{Kotikov}
of the BCDMS and SLAC data, including the higher twists,
                                                    gave smaller values
of $\Lambda_{\overline{\rm MS}}$. It would be interesting to understand
the origin of these smaller values.}
of combined BCDMS and SLAC data on F$_2(x,Q^2)$ structure function:
\begin{equation}
\Lambda^{(4)}_{\overline{\rm MS}}=263 \pm 42 (exp) \ {\rm MeV}
\label{vir}
\end{equation}
As, however, our analysis suggests the importance of including the NNLO
corrections, it would be very desirable to investigate the
                                                        influence of the
yet uncalculated NNLO corrections to F$_2(x,Q^2)$
                                              on the result (\ref{vir})
as well.

So far our analysis has been aimed primarily at the determination of
$\Lambda^{(4)}_{\overline{\rm MS}}$ but it tells us also something
                                                                about
the role of the higher twist contributions to the GLS sum rule.
                                                             Indeed,
Table 1 suggests that if the twist-four contribution (\ref{twist}) is
ignored or its magnitude is negligibly small (like in the bag model
calculations \cite{Fajfer}), the agreement between the results of our
analysis and those of \cite{BCDMS}  worserns. This lends
$a$ $posteriori$ support to the QCD sum rules estimate \cite{Braun}
of the higher twist effects. An independent and more accurate
                                                         estimate of
these higher twists, using for instance lattice methods
                                                 (for a review of such
calculations see \cite{Cris}), would, however, be very welcome.
This analysis might
also shed the light on the role of other possible
                                               1/$Q^2$-contributions,
recently discussed, from a purely theoretical point of view, in
\cite{Zakharov,Altarelli}.
 We intend to consider related problems in the future.

The above results on $\Lambda^{(4)}_{\overline{\rm MS}}$ imply for
 $\alpha_s$ at the scale $\sqrt{Q^{2}}=\sqrt{3}$ GeV the
                                                    following values
\begin{equation}
\alpha_s^{nl}(\sqrt{Q^{2}},\overline{\rm MS}) =
0.318\pm0.010(stat)\pm0.042 (syst) \pm0.026 (twist)\pm0.020 (scheme)
\label{alphanl}
\end{equation}
\begin{equation}
\alpha_s^{nnl}(\sqrt{Q^{2}},\overline{\rm MS}) =
0.315\pm0.010 (stat)\pm0.044 (syst) \pm0.028 (twist)\pm0.005 (scheme)
\label{alphannl}
\end{equation}
where in accordance with the discussions of Sec.2
                                              we have taken $n_{f}=4$.
The errors $\sigma_{\alpha_s}$ of the above results on $\alpha_{s}$ are
related to the errors $\sigma_{\Lambda}$ of (44),(45) as follows
\begin{equation}
\sigma_{\alpha_s}=\frac{\sigma_{\Lambda}}{\Lambda}
\frac{\pi b}{\left|dF/da\right|}
\label{erralph}
\end{equation}
where $F$ is the r.h.s. of eq.(10).

Finally, to facilitate easy comparison with other determinations of
$\alpha_{s}$ we have evolved it, using the fomulae of \cite{Marciano},
and eqs.(\ref{alphanl}),(\ref{alphannl}),
through the b-quark threshold up to $M_{Z}$ and obtained
\begin{equation}
\alpha_s^{nl}(M_Z,\overline{\rm MS})=0.116 \pm 0.001 (stat)\pm0.005
                                                               (syst)
\pm0.003 (twist) \pm0.002 (scheme)
\label{alphas}
\end{equation}
\begin{equation}
\alpha_s^{nnl}(M_Z,\overline{\rm MS})=0.115 \pm 0.001 (stat)
\pm0.005(syst) \pm0.003 (twist) \pm0.0005 (scheme)
\label{alphas2}
\end{equation}

These results can be compared with
\begin{equation}
\alpha_s(M_Z,\overline{\rm MS})=0.113 \pm 0.003(exp) \pm 0.004 (theor)
\label{Mil}
\end{equation}
obtained in the NLO analysis in the $\overline{\rm MS}$ RS
\cite{BCDMS}, where the standard two-loop inverse-log approximation for
$\alpha_s$ was used. The theoretical error comes from the NLO estimates
of the RS and factorization-scheme uncertainties \cite{BCDMS}
 \footnote{We do not use this approximation but are solving
(\ref{bln}) numerically to get $\alpha_{s}$ as a function of the scale
$\mu=\sqrt{Q^{2}}$. Within this procedure, we got at the NLO 0.115
                                                                instead
of
the above mentioned result 0.113. This theoretical difference
                                                      lies within the
total theoretical error of the result (\ref{Mil}).}.

Our results also agree with the recent world average
                                                   \cite{Altarelli}
$$\alpha_s(M_Z,\overline{\rm MS})=0.112\pm 0.007$$
determined on the basis of the results of various
                                            deep-inelastic scattering
experiments. The quoted error bars
correspond to the combination of experimental and theoretical
uncertainties with the dominant role of the latter ones. This average
has in turn been used in the determination of the
                                               overall world average
\cite{Altarelli}
$$\alpha_s(M_Z,\overline{\rm MS})=0.118\pm 0.007$$ with
                                                   the typical error
(mainly theoretical one). This world average sums up all
                                                    available results
on $\alpha_{s}$, including the relatively larger values coming from the
analyses of the R-ratio in the ${\rm e}^+{\rm e}^-$
                                            annihilations into hadrons
and of the total width of $Z^0$ decay into hadrons, both of
which, however, have rather large experimental errors
(see \cite{paper1,Bethke,Branchina,Altarelli}).

{\bf 7.}~~~
In this work we have presented the NNLO QCD analysis
of the GLS sum rules using the new data
 of the CCFR Collaboration \cite{CCFR}. The inclusion of the NNLO
corrections substantially improves the situation as far as the RS
                                                            dependence
is concerned and makes this quantity a potentially good
                                               place for testing the
perturbative QCD. Our result
$$\alpha_s^{nnl}(M_Z,\overline{\rm MS})=
0.115\pm 0.001 (stat) \pm 0.005 (syst) \pm 0.003 (twist) \pm0.0005
(scheme)$$
is in good agreement with the current world average. However, three
 problems, two theoretical and the other experimental,
stand, in the way of a more precise  determination of
$\alpha_s(M_Z)$ or
$\Lambda_{\overline{\rm MS}}$ from this data.
                                         The first one concerns the
treatment of flavour thresholds and we have suggested an approximate
procedure to take this effect into account while still using the
calculations for massless quarks.
The second problem is connected with the necessity of getting more
precise  estimates of the theoretical errors of the higher twist terms.
The third problem is related to the
experimental errors, which are dominated by the systematical ones.
These errors are currently larger then the theoretical errors
of the higher twist terms
and any attempt to reduce them would thus be very usefull.
It would also be very interesting to have accurate data on (\ref{GLS})
                                                                  as a
function of $Q^{2}$ in a broad range of $Q^{2}$ values.
Their analysis would certainly lead to better understanding of the
role of the higher twist contributions to deep inelastic
                                                   scattering at low
and moderate $Q^{2}$.

We intend to carry out an analysis similar to the one presented in
this paper also for the Bjorken sum rule for polarized
                                                  electroproduction
structure functions, since
 the appropriate NNLO perturbative calculations
\cite{LV} as well as
power correction estimates \cite{Balitsky},
                                         inspired by the analysis of
\cite{Anselmino}, are available and the relevant
                                         experimental data will soon
be obtained by CERN and SLAC groups.
\vspace{1.5cm}

\large{\bf Acknowledgements} \normalsize
\vspace{0.5cm}

We are gratefull to the organizers of the XXVII Rencontre de
Moriond, where this work was started, and of the
``QCD-20 Years Later'' Conference, where our preliminary results
were presented,
for the invitations to Les Arcs, France and to Aachen, Germany. We
wish to thank
 J.Blumlein for comments
 on error handling in extraction of $\Lambda_{\overline{\rm MS}}$.
Special thanks are due to G. Altarelli for useful discussions of the
 results of our analysis.
 A. K. wishes to thank the members of the
Theory Division of CERN for their kind hospitality.

\newpage
\parindent 0.0cm
{\bf Figure captions}

\vspace {1.5cm}
Fig.1: The fractions $w^{i},i=1,2,3$ as the functions of $Q^{2}$ for
three
widely used sets of quark distributions functions.

\vspace {1.0cm}
Fig.2: The  dependence of $\Delta(\rho)$ from $n_f$ for the PMS/EC
approaches and the $\overline{\rm MS}$ RS (denoted in this and allthe
following figures by MSb), at the NLO and the NNLO.

\vspace {1.0cm}
Fig.3:  The RS dependence of $\Delta(\rho)$
for $n_{f}=3,4,5$ at the NLO and the NNLO.

\vspace {1.0cm}
Fig.4: The
 comparison between the NLO and the NNLO results for $\Delta(\rho)$
 in the cases of $n_{f}=3,4,5$.

\vspace{1.0 cm}
Note: In the process of the study of Fig.1-Fig.4 the readers
should cnange $\rm MS$ to  $\overline{\rm MS}$ everywhere.
This was not done in view of the technical problems in dealing
with the corresponding PS-file.

\newpage
\begin{table}
\centering
\begin{tabular}{|l|r|r|r|r|r|r|r|r|} \hline
      & RS & $n_{f}$& $r_{0}^{(i)}$ & $r_{1}^{(i)}$ & $r_{2}^{(i)}$ &
$r_{3}^{(i)}$ & $r_{4}^{(i)}$ & $r_{5}^{(i)}$ \\ \hline
    &     & 3 &  0.008137 & 0.645500 &  0.133457 & -0.413737 &
-3.100763 &  7.573763 \\
    & PMS & 4 &  0.008448 & 0.653533 &  0.076082 & -1.053512 &
-3.038260  &  7.176102 \\
    &     & 5 &  0.004056 & 0.814661 & -1.174005 &  0.267708 &
 2.229508  & -2.753092 \\ \cline{2-9}
  N &     & 3 &  0.007837 & 0.653591 &  0.081509 & -0.416852 &
-2.659066  &  7.374106 \\
  N & ECH & 4 &  0.007270 & 0.691163 & -0.264749    & -0.428416    &
-1.482951  &  3.615289 \\
  L &     & 5 &  0.003300 & 0.834586 & -1.291855 &  0.353730 &
2.839872  & -3.593072 \\ \cline{2-9}
    &     & 3 &  0.004927 & 0.760169 & -0.963719    &  0.123290    &
 2.019464  & -2.445137 \\
    & $\overline{\rm MS}$ & 4 & 0.008325 & 0.660719 & -0.062042 &
-1.260045  & -2.644704 & 7.022455 \\
    &     & 5 &  0.008725 & 0.667920 & -0.016647 & -1.436806 &
-2.770146  &  7.534455 \\ \hline
    &     & 3 & 0.004645 & 0.761183 & -0.904944 &  0.435364 & 2.299329
                                                                     &
-3.332708 \\
    & PMS & 4 & 0.008607 & 0.648385 &  0.155498 & -1.095457 & -3.068545
                                                                      &
7.277303 \\
    &     & 5 & 0.006449 & 0.732606 & -0.245552 & -0.864299 & 0.201203
                                                                     &
 2.163065 \\ \cline{2-9}
    &     & 3 & 0.004908 & 0.754859 & -0.903029 &  0.442944 & 1.925275
                                                                     &
-2.829597 \\
  N & ECH & 4 & 0.004350 & 0.782570 & -0.926641 &  0.446002 & 2.117838
                                                                     &
-3.104819 \\
  L &     & 5 & 0.008235 & 0.675962 & 0.148810 & -1.066902 & -3.437890
                                                                     &
7.772996 \\ \cline{2-9}
    &     & 3 & 0.005703 & 0.745372 & -1.420683 & 0.564036 & 3.195462 &
-4.264516 \\
    & $\overline{\rm MS}$ & 4 & 0.002404 & 0.851741 & -2.019862 &
                                                                2.597744
& -0.376554 & -1.574096 \\
    &     & 5 & 0.008854 & 0.668406 & -0.247080 & -1.211435 & -2.402028
                                                                      &
6.958375 \\ \hline
\end{tabular}
\caption{The values
of the coefficients $r^{(i)}_{j}$ for the $i-th$ option
where the option is defined
 by the combination of the order, RS and $n_{f}$.}
\end{table}

\clearpage
\begin{table}
\centering
\begin{tabular}{|l|l|l|l|l|l|l|l|} \hline
       & RS & $n_{f}$ & $\rho$
 & $\Lambda_{\overline{\rm MS}}^{(4)}$ & $\sum_{i=3}^{5}
 w^{i}\Lambda_{\overline{\rm MS}}^{(i)}$ & $\Delta^{(i)}$ &
  $\sum^{5}_{3}w^{(i)} \Delta^{(i)}$ \\ \hline
   &     & 3
& 4.93 & 191 & & 0.170 & \\ \cline{3-5} \cline{7-7}
   & PMS & 4
& 4.65 & 233 & 250$\pm 16(67)$& 0.142 & 0.139 \\ \cline{3-5}
                                                          \cline{7-7}
 N &     & 5
& 4.40 & 321 & & 0.117 & \\ \cline{2-7}
 N &   & 3
& 4.25 & 226 & & 0.151 & \\ \cline{3-5} \cline{7-7}
 L & $\overline{\rm MS}$ & 4
& 4.35 & 250 & 267$\pm 18(79)$ & 0.140 & 0.137 \\ \cline{3-5}
                                                          \cline{7-7}
   &   & 5
& 4.45 & 317 & & 0.121 & \\ \hline
   &  & 3
& 4.50 & 212 & & 0.147 & \\ \cline{3-5} \cline{7-7}
   & PMS & 4
& 4.70 & 230 & 240$\pm 14(60)$ & 0.140 & 0.136 \\ \cline{3-5}
                                                          \cline{7-7}
 N &   & 5 & 5.00 & 278 & & 0.122 & \\ \cline{2-7}
 L &   & 3 & 3.10 & 300 & & 0.139 & \\ \cline{3-5} \cline{7-7}
 & $\overline{\rm MS}$ & 4 & 3.50 & 307 & 318 $\pm 23(99)$ & 0.138 &
0.135 \\ \cline{3-5}
\cline{7-7} &     & 5
& 4.00 & 352 &            & 0.127 &       \\ \hline
\end{tabular}
\caption{Results of the
 procedure described in the text for the case
with twist-four contribution given by the central value of the estimate
(5).
 The values of the $\Lambda$-parameter are in MeV and the first
                                                              (second)
error corresponds to the statistical (systematical) experimental
                                                               errors.}
\end{table}

\begin{table}
\centering
\begin{tabular}{|l|l|l|l|l|l|} \hline
    & RS  & HT=0         & HT=0.016     & HT=0.032     & HT=0.048 \\
                                                                 \hline
NNL & PMS & $333\pm15(65)$ & $294\pm15(65)$ & $250\pm16(67)$ &
$204\pm16(70)$ \\ \cline{2-6}
    & $\overline{\rm MS}$ &
$360\pm19(83)$ & $313\pm19(81)$ & $267\pm18(79)$ & $215\pm17(74)$ \\
                                                               \hline
NL & PMS & $319\pm13(55)$ & $281\pm13(55)$ & $240\pm14(60)$ &
                                                      $201\pm14(60)$
\\ \cline{2-6}
    & $\overline{\rm MS}$ &
$435\pm20(87)$ & $378\pm21(90)$ & $318\pm23(99)$ & $254\pm24(105)$ \\
                                                               \hline
\end{tabular}
\caption{The values of $\Lambda_{\overline{\rm MS}}^{(4)}$ (in MeV)
as extracted from data of the
 CCFR Collaboration under different assumptions of the higher twist
(HT) contribution. The errors are the same as in the Table 2.}
\end{table}


\newpage
\begin{thebibliography}{99}

\bibitem{Altarelli}
G. Altarelli, Talk at the ``QCD-20 Years Later'' Conf. 9-13 June
1992, Aachen; to appear in the Proceedings; preprint CERN-TH.6623/92
(1992).

\bibitem {paper1}
J. Ch\'{y}la, A. L. Kataev and S. A. Larin,
\newblock {\it Phys.Lett.} {\bf B267} (1991) 269.

\bibitem{GKL}
S. G. Gorishny, A. L. Kataev and S. A. Larin,
\newblock in ``Standard Model and Beyond'' Proc. Int. Workshop
CERN-IHEP-JINR, 1--5 October 1990, Dubna, eds. S. Dubnicka, D. Ebert
 and A. Sazonov, World Scientific, Singapore, (1991), p. 288;
{\it Phys. Lett.} {\bf B259} (1991) 144.

\bibitem{Braaten}
E. Braaten, A. Pich and S. Narison,
\newblock {\it Nucl. Phys.} {\bf B373} (1992) 581.

\bibitem{stevenson}
A. C. Mattingly and P. M. Stevenson,
\newblock preprint Rice Univ. DOE/ER/05096-50 (1992).

\bibitem{Deberder}
F. Le Diberder and A. Pich,
\newblock {\it Phys. Lett.} {\bf B286} (1992) 147.

\bibitem{LTV}
S. A. Larin, F. V. Tkachov and J. A. M. Vermaseren,
\newblock {\it Phys. Rev. Lett.} {\bf 66} (1991) 862.

\bibitem{LV}
S. A. Larin and J. A. M. Vermaseren,
\newblock {\it Phys. Lett.} {\bf B259} (1991) 345.

\bibitem{Chet}
K. G. Chetyrkin, S. G. Gorishny, S. A. Larin and F. V. Tkachov,
\newblock {\it Phys. Lett.} {\bf 137B} (1984) 230.

\bibitem{r1MSB}
S. G. Gorishny and S. A. Larin,
\newblock {\it Phys. Lett.} {\bf B172} (1986) 109.


\bibitem {CCFR}
CCFR Collab., M. Shaevitz, Talk at the XXVII Recontre de
Moriond on ``QCD and High Energy Hadronic Interactions'',
22--28 March 1992, Les Arcs; to appear in the Proceedings, ed.
J. Tran Thanh Van;
\\W. C. Leung et al., Nevis Preprint 1460, June 1992.




\bibitem{DO}
D. Duke and J. Owens,
\newblock {\it Phys. Rev.} {\bf D30} (1984) 49.

\bibitem{DFLM}
M. Diemoz, F. Ferroni, E. Longo and G. Martinelli,
\newblock {\it Z. Phys.} {\bf C39} (1988) 21.

\bibitem{MT}
J. Morfin and W. K. Tung,
\newblock {\it Z. Phys.} {\bf C52} (1991) 13.

\bibitem{Jaffe}
R. L. Jaffe and M. Soldate,
\newblock{\it Phys. Rev.} {\bf D26} (1982) 49;
\\E. V. Shuryak and A. I. Vainshtein,
\newblock{\it Nucl. Phys.} {\bf B199} (1982) 951;
\\R. K. Ellis, W. Furmanski and R. Petronzio,
\newblock{\it Nucl. Phys.} {\bf B207} (1982) 1; {\bf B212}
(1983) 29.

\bibitem {Braun}
V. M. Braun and A. V. Kolesnichenko,
\newblock{\it Nucl. Phys.} {\bf B283} (1987) 723.

\bibitem{Zakharov}
V. I. Zakharov,
\newblock Max-Planck-Institut preprint MPI-Ph/92-33 (1992).
\\L. S. Brown, L. G. Yaffe and Chengxing Zhai,
\newblock Univ. of Washington preprint UW/PT-92-07 (1992).



\bibitem{Fajfer}
S. Fajfer and R. J. Oakes,
\newblock{\it Phys. Lett.} {\bf 163B} (1985) 385.


\bibitem {PMS1}
P. M. Stevenson,
\newblock{\it Phys. Rev.} {\bf D23} (1981) 2916.


\bibitem {c2MSB}
O. V. Tarasov, A. A. Vladimirov and A. Yu. Zharkov,
\newblock{\it Phys. Lett.} {\bf 93B} (1980) 429.

\bibitem{Kataev}
A. L. Kataev,
\newblock preprint CERN-TH.6485/92 (1992); in Proc. XXVII
Recontre de Moriond on ``QCD and High Energy Hadronic Interactions''
22--28 March 1992, Les Arcs, ed. J. Tran Thanh Van (to appear).


\bibitem {Grunberg}
G. Grunberg,
\newblock{\it Phys. Lett.} {\bf B221} (1980) 70; preprint
Ecole Polytechnique A510.078 (1982) (unpublished);
{\it Phys. Rev.} {\bf D29} (1984) 2315.

\bibitem{DG}
A. Dhar and V. Gupta,
\newblock{\it Phys. Rev.} {\bf D29} (1984) 2822.

\bibitem{Marciano}
W. Marciano,
\newblock{\it Phys. Rev.} {\bf D29} (1984) 580.


\bibitem{Vovk}
V. I. Vovk,
\newblock{\it Z. Phys.} {\bf C47} (1990) 57.


\bibitem{Kotikov}
A. V. Kotikov, G. Parente and J. Sanchez Guillen,
preprint Univ. of Bern BUTP-91/40 (1991).


\bibitem{BCDMS}
M. Virchaux and A. Milsztajn,
\newblock{\it Phys. Lett.} {\bf B274} (1992) 221.


\bibitem{Cris}
C. Sachrajda, Talk at the ``QCD-20 Years Later'' Conf.
9--13 June 1992, Aachen; to appear in the Proceedings.


\bibitem{Bethke}
S. Bethke and S. Catani,
\newblock preprint CERN-TH.6484/92 (1992);  in Proc.
XXVII Recontre de Moriond ``QCD and High Energy Hadronic Interactions''
22--28 March 1992, Les Arcs, ed. J. Tran Thanh Van (to appear).

\bibitem{Branchina}
V. Branchina, M. Consoli, R. Fiore and D. Zappala,
\newblock{\it Phys. Rev.} {\bf D46} (1992) 75.



\bibitem{Balitsky}
I. I. Balitsky, V. M. Braun and A. V. Kolesnichenko,
\newblock{\it Phys. Lett.} {\bf B242} (1990) 245.


\bibitem{Anselmino}
M.Anselmino, B.L.Ioffe and E.Leader,
\newblock{\it Yad.Fyz.} {\bf 49} (1989) 214.


\end{thebibliography}
\end{document}